\documentclass{emulateapj}
\usepackage{amsmath}
\usepackage{multirow}
\usepackage{array}
\usepackage[dvipsnames]{xcolor}
\usepackage{hyperref}

\newcommand{\code}[1]{\texttt{#1}}
\newcommand{\mesa}{\code{MESA}}
\newcommand\logten{\ensuremath{\log_{10}}}
\newcolumntype{C}[1]{>{\centering\let\newline\\\arraybackslash\hspace{0pt}}m{#1}}

\shorttitle{SN 2016gkg Progenitor Constraints Using Statistical Analysis}
\shortauthors{Sravan et al.}
\begin{document}

%\title{Progenitors and Binary Companions of SN 2016gkg}
\title{Constraints on the Progenitor System of SN 2016gkg \\ from a Comprehensive Statistical Analysis}

\author{Niharika Sravan, Pablo Marchant, Vassiliki Kalogera, and Raffaella Margutti}
\affil{Center for Interdisciplinary Exploration and Research in Astrophysics (CIERA) \\
and \\
Department of Physics and Astronomy \\
Northwestern University, 2145 Sheridan Road, Evanston, IL 60208, USA}

%\collaboration{(AAS Journals Data Scientists collaboration)}
%\nocollaboration




% ApJL Restrictions:
% Abstract: no more than 250 words
% Main text: no more than 3500 words
% Figures and tables: no more than five in any combination, e.g., three figures and two tables. Tables longer than 50 rows will automatically be converted to machine-readable table format and published online only. Note that there is a limit of one machine-readable table per manuscript. Multi-panel figures are limited to nine panels per figure.
% References: no more than 50 references

\begin{abstract}
Type IIb supernovae (SNe) present a unique opportunity for understanding the progenitors of stripped-envelope (SE) SNe as the stellar progenitor of several Type IIb SNe have been identified in pre-explosion images. In this paper, we use Bayesian inference and a large grid of non-rotating solar-metallicity single and binary stellar models to derive the associated probability distributions of single and binary progenitors of the Type IIb SN 2016gkg using existing observational constraints. We find that potential binary star progenitors have smaller pre-SN hydrogen-envelope and helium-core masses than potential single-star progenitors typically by $0.1M_\odot$ and $2M_\odot$, respectively. We find that, a binary companion, if present, is a main-sequence or red-giant star. Apart from this, we do not find strong constraints on the nature of the companion star. We demonstrate that the range of progenitor helium-core mass inferred from observations could help improve constraints on the progenitor. We find that the probability that the progenitor of SN 2016gkg was a binary is 22\% when we use constraints only on the progenitor luminosity and effective temperature. Imposing the range of pre-SN progenitor hydrogen-envelope mass and radius inferred from SN light-curves the probability the progenitor is a binary increases to 44\%. However, there is no clear preference for a binary progenitor. This is in contrast to binaries being the currently favored formation channel for Type IIb SNe. Our analysis demonstrates the importance of statistical inference methods to constrain progenitor channels.
\end{abstract}

\keywords{binaries: general -- stars: massive -- supernovae: general -- supernovae: individual (SN 2016gkg)}

\section{Introduction} \label{s:intro}

The mechanisms driving the stripping of the progenitor stars of stripped-envelope (SE) supernovae (SNe) remain an open research question. 
Currently, close binary interactions, stellar winds, and nuclear burning instabilities are leading candidates to explain the mass loss \citep[e.g.,][]{2011A&A...528A.131C,2011ApJ...741...33A,2013A&A...558A.131G,2014ARA&A..52..487S,2017MNRAS.470L.102S}.  
Among SE SNe, Type IIb SNe explode with a low-mass residual hydrogen-envelope \citep[e.g.,][]{1992ApJ...391..246P, 1994ApJ...429..300W} and initially exhibit prominent hydrogen spectral features that weaken and disappear in the weeks following the explosion.
The progenitors of five Type IIb SNe have been identified in pre-explosion images: 1993J \citep{1994AJ....107..662A}, 
%2001ig \citep{2006MNRAS.369L..32R}, 
2008ax \citep{2015ApJ...811..147F}, 2011dh \citep{2011ApJ...739L..37M,2011ApJ...741L..28V}, 2013df \citep{2014AJ....147...37V}, and 2016gkg \citep{2017MNRAS.465.4650K}. Furthermore, there is evidence for the presence of binary companions to the progenitors of SNe 1993J and 2011dh \citep{2014ApJ...790...17F,2014ApJ...793L..22F}. This makes Type IIb SNe ideal candidates to test theories of binary evolution.

SN 2016gkg was discovered on 2016 September 20.18 UT in NGC 613. %reported by A. Buso and S. Otero https://wis-tns.weizmann.ac.il/object/2016gkg. 
%Broad H$\alpha$ emission and P-Cygni features were detected in its spectrum on 2016 September 25.33 UT. Broad He I features appeared on 2016 28.56 September UT. 
%\cite{2017MNRAS.465.4650K} identify a source in pre-explosion Hubble Space Telescope (HST) images as its progenitor and infer its luminosity and effective temperature (see Table \ref{t:prog_prop}). However, \cite{2017ApJ...836L..12T}, using the same images, identify two sources at the SN location and conclude that they cannot favor one over the other as the progenitor or rule out that they are star clusters. \cite{2017ApJ...837L...2A} fit analytic models to the light curve of SN 2016gkg and derive a radius and residual hydrogen-envelope mass for the progenitor star (see Table \ref{t:prog_prop}).
\cite{2017MNRAS.465.4650K} identified a source in pre-explosion Hubble Space Telescope (HST) images as its progenitor and inferred its luminosity and effective temperature. \cite{2017ApJ...836L..12T} found an additional source and concluded they could not favor either source as the progenitor star. They also found different magnitudes for the common source. The properties for this source inferred by \cite{2017ApJ...836L..12T} are consistent (within 1 and $3\sigma$ in luminosity and effective temperature, respectively) with those inferred by \cite{2017MNRAS.465.4650K}. Therefore, for simplicity, we adopt the constraints of \citet{2017MNRAS.465.4650K}.
\cite{2017ApJ...837L...2A} fit analytic models to the light curve of SN 2016gkg and derived a radius and residual hydrogen-envelope mass for the progenitor star (see Table \ref{t:prog_prop}).

\begin{deluxetable*}{cccc}
\tablewidth{0pt}
\tabletypesize{\footnotesize}
\tablecaption{{{Properties of Detected SN 2016gkg Progenitor}} \label{t:prog_prop}}
\tablehead{\colhead{$\logten (L/L_\sun)$} & \colhead{$T_{\rm eff}/$K} & \colhead{$R/R_\sun$} & \colhead{$M_{\rm env}/M_\sun$}}
\startdata
%1993J & $5.1 \pm 0.3$ $^{a*}$  & $3.63 \pm 0.05$ $^{a*}$ & \nodata & \nodata & $575 \pm 144$ $^b$ & $0.15 - 0.3$ $^{a,b}$ & $3.5 - 5.1$ $^{a,b}$ & $13 - 17$ $^{a,c}$ & $5.0 \pm 0.3$ $^{a*}$ &  $4.3 \pm 0.1$ $^{a*}$ & 14 $^a$ \\
%2001ig & $4.9 - 5.3$ $^d$& \nodata & \nodata & \nodata & \nodata & \nodata &  & $\sim 15$ $^e$ & $\sim 4.5$ & \nodata & \nodata \\ 
%2008ax & $4.42 - 5.3$ $^f$ & $3.88 - 4.3$ $^f$ & \nodata & \nodata & $30 - 50$ $^f$ & $0.01- 0.1$ $^{g,h}$ & $3 - 6$ $^g$ & $10 - 14$ $^{g,i}$ & \nodata & \nodata & 12 $^f$ \\ 
%2011dh & $4.92 \pm 0.20$ $^j$ & $3.78 \pm 0.02$ $^j$ & 21.808 $^k$ & 21.208 $^k$ & $200 - 300$ $^l$ & $0.01 - 0.1$ $^{l,m}$ & $2.7 - 4$ $^{l,n}$ & $10 - 15$ $^{l,m,o}$ & \nodata & \nodata & \nodata  \\
%2013df & $4.94 \pm 0.06$ $^p$ & $3.63 \pm 0.01$ $^p$ & 24.515 $^p$ & 23.106 $^p$ & $545 \pm 65$ $^p$ & $\sim 0.2$ $^q$ & $2 - 3.6$ $^q$ & $12 - 17$ $^{p,q}$ & \nodata & \nodata & \nodata \\
\\
%2016gkg & 
%5.14$^{+0.22}_{-0.39}$ $^a$ & 9500$^{+6100}_{-2900}$ $^a$ & & & 40 -- 150 $^b$ & 0.02 -- 0.4 $^b$ & \nodata & & & & \\
5.14$^{+0.36}_{-0.14}$ $^a$ & 9500$^{+3700}_{-1033}$ $^a$ & 40 -- 150 $^b$ & 0.02 -- 0.4 $^b$ \\
\enddata
%\tablecomments{{\bf (1)} $L_{1}$: Progenitor luminosity
%\\ {\bf (2)} $T_{\rm eff,1}$: Progenitor effective temperature
%\\ {\bf (3)} $V_1$: Progenitor V-band magnitude
%\\ {\bf (4)} $I_1$: Progenitor I-band magnitude
%\\ {\bf (3)} $R_1$: Progenitor radius
%\\ {\bf (4)} $M_{\rm env,1}$: Progenitor Hydrogen envelope mass
%\\ {\bf (5)} $M_{\rm core,1}$: Progenitor Helium core mass
%\\ {\bf (8)} $M_{\rm ZAMS,1}$: Progenitor ZAMS mass
%\\ {\bf (9)} $L_{2}$: Companion luminosity
%\\ {\bf (10)} $T_{\rm eff,2}$: Companion effective temperature
%\\ {\bf (11)} $M_{\rm ZAMS,2}$: Companion ZAMS mass
%}
%\tablenotetext{a}{}
\tablerefs{$^a$ Kilpatrick, C. D., private communication (uncertainties are one-third of $3\sigma$, see Section \ref{s:sm}); $^b$ \cite{2017ApJ...837L...2A}.}
\end{deluxetable*}

In this paper, given observational constraints on its progenitor properties, we use Bayesian inference to derive the distribution of properties of potential progenitor systems (both singles and binaries) of SN 2016gkg. We also calculate the probability that the progenitor was a binary. We assume that the constraints derived by \citet{2017MNRAS.465.4650K} corresponds to the progenitor. We discuss the effect of using the pre-SN progenitor hydrogen-envelope and helium-core mass constraints to distinguish between single and binary progenitor channels.

This paper is organized as follows. In Section \ref{s:method} we briefly describe our models and method. In Section \ref{s:results} we discuss results for the distribution of potential progenitor systems (both singles and binaries) of SN 2016gkg. We summarize our results and conclude in Section \ref{s:conclusions}.

\section{Method} \label{s:method}

\subsection{Single and Binary Star Models} \label{s:models}

We compute a large grid of non-rotating solar-metallicity\footnote{We choose the value of solar metallicity ($Z_\sun$) to be 0.02. The metallicity of the host galaxy of SN 2016gkg is $0.012\pm0.004$ \citep{2017MNRAS.465.4650K}. See Conclusions for a discussion on the effect of metallicity.} single and binary star models with Modules for Experiments in Stellar Astrophysics \citep[\mesa\footnote{Release 9575.},][]{2011ApJS..192....3P,2013ApJS..208....4P,2015ApJS..220...15P}. We briefly summarize the models in what follows. 

We start the evolution of the star(s) at the zero-age main sequence (ZAMS). We stop the evolution if any of the following conditions are met: the carbon mass fraction at (any) star's center is lower than $10^{-6}$, the hydrogen-envelope mass of any star drops below $0.01M_\sun$ (in which case we assume the system is a completely stripped Type Ibc progenitor), or, in binaries, the accretor overfills its Roche-lobe. We assume the surface properties of the star at carbon depletion match those of the pre-supernova progenitor star. This is because the thermal timescale of the envelope is large compared to the time between carbon depletion and iron core-collapse. We note however that it has recently been proposed that waves could efficiently transport energy outwards during core neon and oxygen burning, potentially producing outbursts and large changes in the progenitor surface luminosity and effective temperature months or years prior to the explosion \citep{2012MNRAS.423L..92Q,2014ApJ...780...96S,2017MNRAS.470.1642F}.

We use the \texttt{basic.net}, \texttt{approx21.net}, and \texttt{co\_burn} \texttt{.net} nuclear networks in \mesa. We adopt the standard mixing-length theory and the Ledoux criterion to model convection, with $\alpha_{\rm MLT}$ set to 1.5. When convective regions approach the Eddington limit, the efficiency of convection is enhanced\footnote{The treatment of these regions is a subject of debate and stellar evolution calculations during these phases are uncertain.} \citep{2013ApJS..208....4P}. 
To account for the nonzero momentum of a convective element at the Hydrogen burning convective core boundary, we extend this region by 0.335 of the pressure scale height \citep{2011A&A...530A.115B}. We adopt the value of dimensionless free parameter for semi-convection, $\alpha_{\rm sc}$, to be 1.0 \citep{2006A&A...460..199Y}. We use radiative opacity tables from the OPAL project \citep{1996ApJ...464..943I}. We adopt surface effective temperature and abundance dependent stellar wind prescriptions. When $T_{\rm eff} > 10^4\,$K, we adopt the prescription of \citet{2001A&A...369..574V} if the surface hydrogen mass fraction $>$ 0.4 and \citet{2000A&A...360..227N} otherwise. If $T_{\rm eff} < 10^4\,$K we adopt the prescription of \citet{1988A&AS...72..259D}. 

We use %the `implicit' scheme and 
the model of \citet{1990A&A...236..385K} to calculate the mass transfer rate due to Roche-lobe overflow (RLO) in our binary star models. The efficiency of mass transfer (the ratio of mass accreted by the secondary to the mass transferred via RLO by the primary), $\epsilon$, is assumed to be constant during the evolution. 
The mass not accreted is assumed to be lost as stellar winds. Stellar winds carry away with them the specific angular momentum of the corresponding component. All orbits are assumed to be circular.

We compute single-star models with initial mass, $\logten(M_{\rm ZAMS}/M_\sun)$ = 1.28 -- 1.40 ($M_{\rm ZAMS}/M_\sun \sim$ 19 -- 25) in intervals of 0.0005 dex and binary star models with initial primary mass, $\logten(M_{\rm ZAMS,1}/M_\sun)$ = 1.0 -- 1.4 ($M_{\rm ZAMS,1}/M_\sun \sim$ 10 -- 25) in intervals of 0.02 dex, initial mass ratio, $q_{\rm ZAMS} \equiv M_{\rm ZAMS,2}/M_{\rm ZAMS,1}$ =  0.225 -- 0.975 in intervals of 0.05, and initial orbital period, $\logten(P_{\rm orb}/$d$)$ = 2.5 -- 3.8 ($P_{\rm orb}/$d$ \sim$ 316 -- 6310) in intervals of 0.02 dex\footnote{\mesa\ inlists used for these can be found at \url{https://github.com/orlox/mesa_input_data/tree/master/2017_IIb}}. We choose this parameter space based on a broader scan. We compute the models for $\epsilon$ = 0.5 and 0.1. Models that reach core carbon exhaustion ($C_{\rm center} \leq 10^{-6}$) with less than $1M_\sun$ (but greater than $0.01M_\sun$) of residual hydrogen-envelope are defined as Type IIb SN progenitors. This criterion is a conservative choice as residual hydrogen envelope masses of Type IIb SNe with detected progenitors are less than $0.5 M_\odot$ \citep{1994ApJ...429..300W}. Type IIb SNe with detected progenitors represent those with the most massive envelopes: progenitors with smaller envelopes are compact \citep{2017ApJ...840...10Y} and thus harder to detect. Moreover, the cooling envelope feature in Type IIb SN light curves decrease with decreasing radius \citep{2016MNRAS.455..423M} and envelope mass making compact Type IIb SNe harder to detect.

%\begin{deluxetable}{lccc}
%\tabletypesize{\scriptsize}
%\tablecaption{{{Binary Grid}} \label{t:bgrid}}
%\tablewidth{0pt}
%\tablehead{\colhead{\ } & \colhead{min} & \colhead{max} & \colhead{interval}}
%\startdata
%$M_1/M_\sun$ & 10 & 30 & 2 \\
%$M_2/M_\sun$ & 6 & $M_1$ & 2 \\
%$\logten(P_{\rm orb}$/days) & 0 & 4 & 0.1 \\
%\enddata
%\tablenotetext{}{{\bf (1)} $M_1$: Initial Primary Mass}
%\tablenotetext{}{{\bf (2)} $M_2$: Initial Secondary Mass}
%\tablenotetext{}{{\bf (3)} $P_{\rm orb}$: Initial Orbital Period}
%\end{deluxetable}

\subsection{Statistical Method} \label{s:sm}

We use Bayesian inference to derive the distribution of the potential progenitors (and their binary companions) of SN 2016gkg. We adopt $3\sigma$ (see below for reasoning) luminosity and effective temperature constraints on the progenitor as derived from observations of the progenitor system before explosion %inferred by \cite{2017MNRAS.465.4650K} 
(Kilpatrick, C. D., private communication). We assume that luminosity and effective temperature are independent variables for simplicity, though this assumption is not accurate. We do not apply the progenitor hydrogen-envelope mass and radius constraints derived from the SN light-curves as these are model-dependent. However, we discuss the implications of applying them later. 

\begin{figure*}[t!]
\begin{center}
\includegraphics[width=\textwidth]{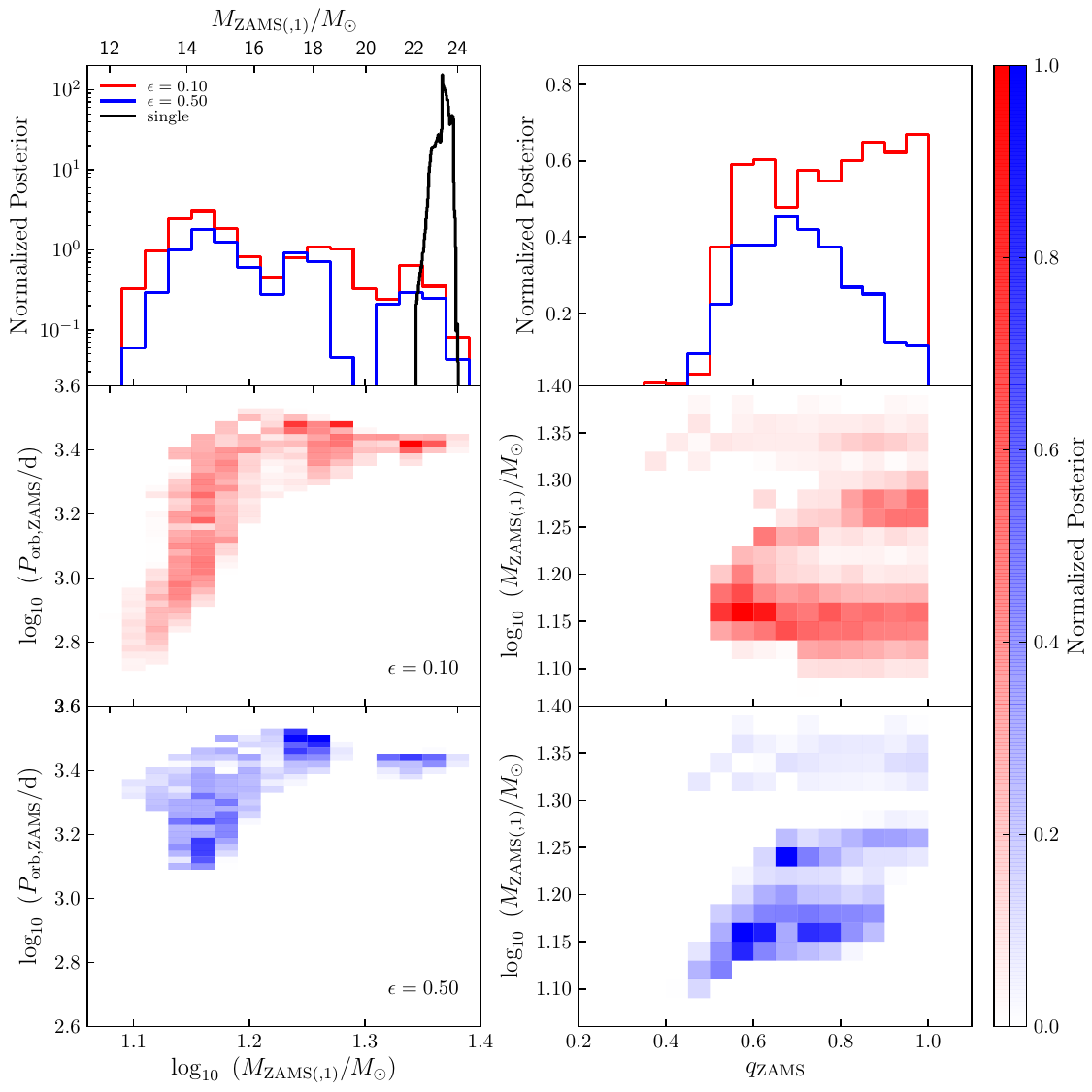}
\end{center}
\caption{Posterior probability distributions of the parameter space of potential single and binary star progenitors of SN 2016gkg. The top panel shows the distribution of initial (primary) mass ($M_{\rm ZAMS(,1)}$, left) and, for binaries, initial mass ratio ($q_{\rm ZAMS} \equiv M_{\rm ZAMS,2}/M_{\rm ZAMS,1}$, right) of potential single (black) and binary star progenitors with $\epsilon$ = 0.1 (red) and 0.5 (blue). The histograms show the total posterior probability in each bin. The histograms in the top left panel have been rescaled such that areas under them reflect the probability of a single or binary star progenitor of SN 2016gkg.
The middle and bottoms panels show 2-D distributions of $P_{\rm orb}$ and $M_{\rm ZAMS(,1)}$ (left) and $M_{\rm ZAMS(,1)}$ and $q_{\rm ZAMS}$ (right) of potential binary star progenitors with $\epsilon$ = 0.1 (middle) and 0.5 (bottom). 
\label{f:fig1}}
\end{figure*}

\begin{figure*}
\begin{center}
\includegraphics[width=\textwidth]{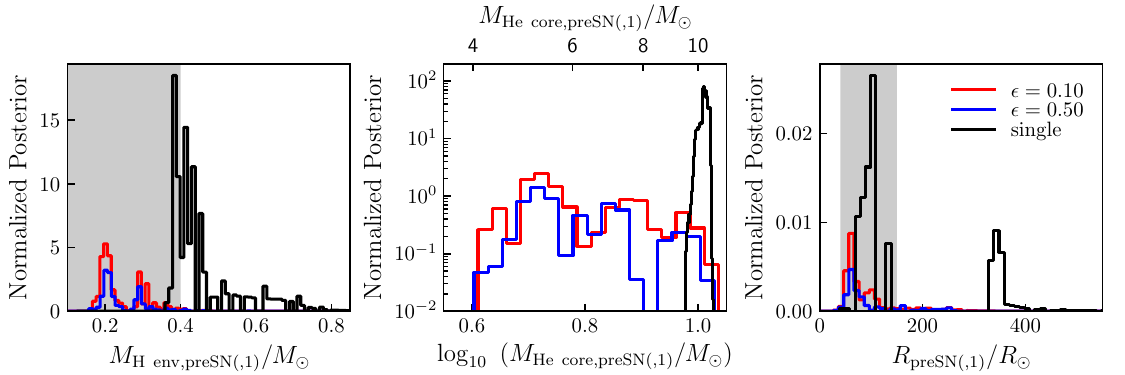}
\end{center}
\caption{Distributions of pre-SN properties of potential single (black) and binary ($\epsilon$ = 0.1 (red) and 0.5 (blue)) star progenitors of SN 2016gkg: progenitor hydrogen-envelope mass ($M_{\rm H~env,preSN(,1)}$, left), progenitor helium-core mass ($M_{\rm He~core,preSN(,1)}$, middle), and progenitor radius ($R_{\rm preSN(,1)}$, right). The histograms show the total posterior probability in each bin and have been rescaled such that areas under them reflect the probability of a single or binary star progenitor of SN 2016gkg. Grey shaded regions indicate constraints derived from light-curves (see Table \ref{t:prog_prop}). Pre-SN hydrogen-envelope and helium-core mass for potential binary star progenitors are smaller than for potential single-star progenitors typically by $0.1M_\odot$ and $2M_\odot$, respectively. Constraints on the progenitor hydrogen-envelope mass and radius from light-curves increase the likelihood of a binary progenitor of SN 2016gkg from 22\% (13\%) to 44\% (28\%) for $\epsilon$ =  0.1 (0.5).
\label{f:fig2}}
\end{figure*}

\begin{figure}
\begin{center}
\includegraphics{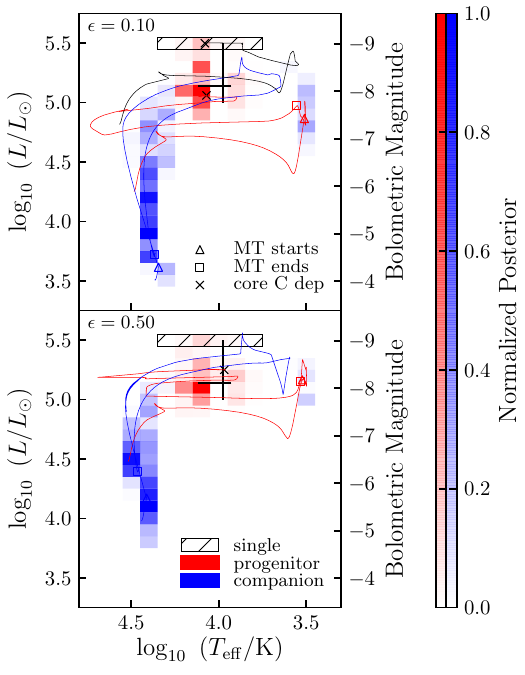}
\end{center}
\caption{Distributions of H-R locations of potential single and binary ($\epsilon$ = 0.1 (top) and 0.5 (bottom)) star progenitors of SN 2016gkg. The red (blue) color scales show distributions for the primary (secondary) of binary star progenitors. Red (blue) H-R tracks are for the primary (secondary) of the binary star model with the highest posterior probability for the corresponding $\epsilon$: $M_{\rm ZAMS,1}$ = 14.45 (17.38) $M_\sun$, $q_{\rm ZAMS} \equiv M_{\rm ZAMS,2}/M_{\rm ZAMS,1}$ = 0.575 (0.675), and $P_{\rm orb}$ = 2291 (3311) days for $\epsilon$ = 0.1 (0.5). Mass transfer (MT) is defined to be taking place when mass transfer rate due to RLO is $\geq 10^{-6} M_\sun$ yr$^{-1}$. The black H-R track is for the single-star model with the highest posterior probability: $M_{\rm ZAMS}$ = 23.28 $M_\sun$. The hatched region shows H-R locations of potential single-star progenitors with normalized posterior probabilities $\geq 0.01$. Black error-hairs show one-third of the $3\sigma$ constraints on the observed progenitor for SN 2016gkg (Table \ref{t:prog_prop}).
%Left to right: Distributions of H-R locations and V, R, and I color magnitudes of potential single and binary star progenitors of SN 2016gkg. The top (bottom) H-R diagram shows binary star progenitors with $\epsilon$ = 0.1 (0.5). The top (bottom) panels of distributions of color magnitudes are those for the primary (companion) and red (blue) lines show distributions for binary star progenitors with $\epsilon$ = 0.1 (0.5). Black lines show distributions for single-star progenitors. The distributions in all panels have been normalized to the maximum posterior probability in their bins.
If present, the binary companion of SN 2016gkg has $M_{\rm bol} \gtrsim -8.5$ and is a main-sequence or red-giant star (for binary star progenitors with initial mass ratios $\sim 1$). As such no strong constraints can be placed on the companion.
\label{f:fig3}}
\end{figure}

%For each Type IIb SN progenitor model (see above for definition) we calculate an unnormalized posterior probability ($P'$):
%\begin{equation} \label{e:post}
%P(\bar{\theta}_{\rm mod}|\bar{X}_{\rm obs}) = \mathcal{L}(\bar{X}_{\rm obs}|\bar{X}_{\rm mod}) P(\bar{\theta}_{\rm mod})
%\end{equation}
%\begin{equation}
%P'(\bar{X}_{\rm obs}|\bar{X}_{\rm mod}) = P(\bar{\theta}_{\rm obs}|\bar{X}_{\rm mod})
%\end{equation}
For each Type IIb SN progenitor model (see above for definition) we compute a posterior probability ($P$):
\begin{equation} \label{e:post}
P(\vec{\theta}_{\rm mod}|\vec{X}_{\rm obs}) = \mathcal{L}(\vec{X}_{\rm obs}|\vec{\theta}_{\rm mod}) P(\vec{\theta}_{\rm mod}), 
\end{equation}
modulo the standard normalization constant in Bayes' theorem. 
The model parameters, $\vec{\theta}_{\rm mod}$, are the initial single (initial mass) or binary (initial primary mass, mass ratio, and orbital period) star progenitor properties and eventually determine the pre-SN progenitor properties, $\vec{X}_{\rm mod}$, i.e., $\vec{X}_{\rm mod}(\vec{\theta}_{\rm mod})$.
%\begin{equation}
%P'(\bar{X}_{\rm mod}|\bar{X}_{\rm obs}) =  \mathcal{L}(\bar{X}_{\rm obs}|\bar{X}_{\rm mod}) P_0(\bar{X}_{\rm mod})
%\end{equation}
%\begin{equation}
%P'(\bar{X}_{\rm mod}|\bar{X}_{\rm obs}) = P(\bar{\theta}_{\rm mod}|\bar{X}_{\rm obs})
%\end{equation}
%where, $\bar{X}_{\rm obs}$ is the vector of observed luminosity and effective temperature of the progenitor of SN 2016gkg, $\bar{X}_{\rm mod}$ is the vector of luminosity and effective temperature of the model progenitor, and $\bar{\theta}_{\rm mod}$ is the vector of model parameters. We assume that 
%\begin{equation}
%P(\bar{X}_{\rm mod}) = P(\bar{\theta})
%\end{equation}
%for $\bar{\theta} \in [\bar{\theta}_{\rm mod}-\Delta\bar{\theta}_{\rm mod}/2,\bar{\theta}_{\rm mod}+\Delta\bar{\theta}_{\rm mod}/2]$, where $\Delta\bar{\theta}_{\rm mod}$ is the vector of spacing in the parameter space scan for $\bar{\theta}_{\rm mod}$.
%We adopt the split normal distribution\footnote{$\mathcal{L}(\mu,\sigma_+,\sigma_-|x) = \sqrt{\frac{2}{\pi}} \frac{1}{\sigma_- +\sigma_+} \exp\left(-\frac{(x-\mu)^2}{2\sigma^2}\right)$, $\sigma$ = $\sigma_-$ when $x<\mu$ and $\sigma$ = $\sigma_+$ when $x\geq\mu$.} for $\mathcal{L}(\bar{X}_{\rm obs}|\bar{X}_{\rm mod})$ and one-third of the $3\sigma$ constraints on the progenitor luminosity and effective temperature (Kilpatrick, C. D., private communication) as $1\sigma$ constraints (Table \ref{t:prog_prop}).
For each individual observable quantity $X_{\rm obs,i}$ (the $i$-th component of vector $\vec{X}_{\rm obs}$) with mean, $\mu_{\rm i}$, and uncertainties, $\sigma_{\rm +/-,i}$, we adopt the split normal distribution for the likelihood
\begin{equation}
\begin{split}
\mathcal{L}(\mu_{\rm i},\sigma_{\rm +,i},\sigma_{\rm -,i}|X_{\rm mod,i}) = & \sqrt{\frac{2}{\pi}} \frac{1}{\sigma_{\rm -,i} +\sigma_{\rm +,i}} \times \\
												&\exp\left(-\frac{(X_{\rm mod,i}-\mu_{\rm i})^2}{2 \sigma^2_{\rm i} } \right)
\end{split}
\end{equation} 
where $\sigma_{\rm i}$ = $\sigma_{\rm -,i}$ when $X_{\rm mod,i}<\mu_{\rm i}$ and $\sigma$ = $\sigma_{\rm +,i}$ when $X_{\rm mod,i}\geq\mu_{\rm i}$. 
For each individual observable quantity $X_{\rm obs,i}$ with a range of values, we adopt a flat distribution for the likelihood. Thus,
\begin{equation}
\mathcal{L}(\vec{X}_{\rm obs}|\vec{\theta}_{\rm mod}) = \prod_{\rm i} \mathcal{L}(\mu_{\rm i},\sigma_{\rm +,i},\sigma_{\rm -,i}|X_{\rm mod,i})
\end{equation} 
% for $\mathcal{L}(\bar{X}_{\rm obs}|\bar{X}_{\rm mod})$ and one-third of the $3\sigma$ constraints on the progenitor luminosity and effective temperature (Kilpatrick, C. D., private communication) {\bf as} $1\sigma$ constraints (Table \ref{t:prog_prop}).

Observational uncertainties are not necessarily distributed as a normal distribution and this form of the likelihood function is just an approximation. In particular, the $1\sigma$ range for luminosity and effective temperature from \citet{2017MNRAS.465.4650K} is much wider than one-third of their $3\sigma$ range (Kilpatrick, C. D., private communication). To avoid artificially extending the range of uncertainty in the observations, we adopt $\sigma_{\rm +}$ and $\sigma_{\rm -}$ to be one-third of the respective $3\sigma_{\rm +}$ and $3\sigma_{\rm -}$ values instead of the $1\sigma_{\rm +/-}$ values in \citet{2017MNRAS.465.4650K}.
%We use one-third of $3\sigma$ instead of $1\sigma$ constraints from \citet{2017MNRAS.465.4650K} to obtain the true likelihood of hot stars. Low (high) likelihood of hot stars would tighten (extend) the ZAMS mass range for single but not binary star progenitors. This is because the effective temperature of single star progenitors is a function of its ZAMS mass due to scaling in mass-loss (binary star progenitors can adopt a range of effective temperatures at the same ZAMS mass depending on when RLO begins).

The prior probability $P(\vec{\theta}_{\rm mod})$ is computed for the range $[\vec{\theta}_{\rm mod}-\Delta\vec{\theta}_{\rm mod}/2,\vec{\theta}_{\rm mod}+\Delta\vec{\theta}_{\rm mod}/2]$. For a single-star with initial mass, $\logten M_{\rm ZAMS}$,
\begin{equation}
P(\vec{\theta}_{\rm mod}) = (1-f_{\rm bin})P(\logten M_{\rm ZAMS})
\end{equation}
and for a stellar binary with initial primary mass, $\logten M_{\rm ZAMS,1}$, initial mass ratio, $q_{\rm ZAMS}$, and initial orbital period, $P_{\rm orb}$,
\begin{equation}
\begin{split}
P(\vec{\theta}_{\rm mod}) = f_{\rm bin}P(\logten M_{\rm ZAMS,1})&P(q_{\rm ZAMS})\\
					&P(\logten P_{\rm orb})
\end{split}
\end{equation}
where, $f_{\rm bin}$ is the fraction of stars in binaries.

We assume $f_{\rm bin}$ to be a constant and independent of $M$, $q$, and $P_{\rm orb}$. The distribution of $M$ is taken to be the Salpeter Initial Mass Function \citep[IMF,][]{1955ApJ...121..161S}
\begin{equation} \label{e:IMF}
f(M) = M^{-\alpha}; 
\end{equation}
We assume that the minimum ZAMS mass needed to undergo core-collapse is $8 M_\sun$ \citep{2002RvMP...74.1015W}. We adopt a power-law distribution for the initial mass ratio, $q$, 
\begin{equation} \label{e:f(q)}
f(q) = q^\beta
\end{equation}
This distribution is assumed to be followed for $q>0.2$ \citep{2014ApJS..213...34K}.
Finally, the distribution of initial orbital period, $P_{\rm orb}$, is chosen according to \citet{2014ApJS..213...34K}
\begin{equation} \label{e:f(P)}
f(\logten P_{\rm orb})=(\logten P_{\rm orb})^{\gamma}
\end{equation}
This distribution is assumed to hold\footnote{The upper limit for the validity of this distribution is 2000 days \citep{2014ApJS..213...34K}. However, due to poor constraints for wide binaries we assume that this distribution holds up to 10,000 days.} for $0.15 < \logten (P_{\rm orb}/$d$) < 4$.

%Using Eqs. \ref{e:IMF} and \ref{e:f(q)}, $P(M_1,M_2)$ is given by:
%
%\begin{equation}
%P(M_1,M_2) = \frac{
%\int\limits_{M_1-(\Delta M_1/2)}^{M_1+(\Delta M_1/2)}  dM^\prime_1 M_1^{\prime-(\alpha+\gamma+1)}
%\int\limits_{M_2-(\Delta M_2/2)}^{M_2+(\Delta M_2/2)} dM^\prime_2 M_2^\gamma
%}
%{
%\int\limits_{8}^{\infty} dM^\prime_1 \int\limits_{1.5}^{M^\prime_1-(\Delta M_1/2)} dM^\prime_2~M_1^{\prime-(\alpha+\gamma+1)} M_2^{\prime\gamma}
%}
%\end{equation}

\section{Results} \label{s:results}

We compute posterior probabilities for SN 2016gkg using our model Type IIb SN progenitors (see Section \ref{s:models} for definition) using the method described above. Unless otherwise mentioned, we assume $f_{\rm bin} = 0.5$, $\alpha = 2.3$ \citep{1955ApJ...121..161S}, $\beta$ = -1, and $\gamma = -0.22$ \citep{2014ApJS..213...34K}. 

Some binary star models experience very little interaction, transferring only small amounts of mass when the primary star's atmosphere Roche-lobe overflows. Therefore their evolution largely resembles that of single stars. We therefore require that primaries transfer at least 1\% of their initial mass in RLO to qualify as `binary' progenitors. The exact choice for this number does not affect our results significantly; lowering it by an order of magnitude adds some $\gtrsim 22 M_\odot$ mass binaries with net posterior probabilities $\lesssim$3\% more for our fiducial priors.

In figure \ref{f:fig1} we show the distribution of the parameter space of potential single and binary star progenitors of SN 2016gkg. There are three peaks in the distribution of initial primary mass for binary star progenitors. The low mass peak is favored by the prior on initial primary mass (Eq. \ref{e:IMF}), the middle peak is due to the likelihood for $T_{\rm eff}$, and the high mass peak results from mildly interacting binaries with relatively undisturbed primaries whose evolution largely resembles their single-star counterparts. There are fewer binary star progenitors with $\epsilon = 0.5$, $q_{\rm ZAMS}>0.7$, and $\logten(P_{\rm orb}/$d$) \lesssim 3.1$ as they experience unstable mass transfer or evolve into contact, which lead to a merger.%in which case we assume a common envelope develops. %Models run without MLT++ favor slightly larger initial primary masses: the dominant peak is at $\logten(M_{\rm ZAMS,1}/M_\sun) \sim 1.2$. However, the distributions of initial mass ratio and initial orbital period are robust to the effects of MLT++.

In figure \ref{f:fig2} we show the distribution of pre-SN properties of potential single and binary star progenitors of SN 2016gkg. The three peaks in the distributions of initial primary mass (Figure \ref{f:fig1}) of binary star progenitors roughly translate to the distributions of pre-SN hydrogen-envelope and helium-core mass. Pre-SN hydrogen-envelope and helium-core mass for potential binary star progenitors are clearly smaller than for potential single-star progenitors \citep[e.g.,][]{1992ApJ...391..246P, 2010ApJ...725..940Y, 2011A&A...528A.131C} typically by $0.1M_\odot$ and $2M_\odot$, respectively. Therefore, these can be used to distinguish progenitor scenarios. While progenitor helium-core mass constraints are currently unavailable for SN 2016gkg, their existence could increase the likelihood of a binary progenitor of SN 2016gkg significantly by ruling out several single-star progenitors (see below for a discussion on rates). The distribution of all binary star properties shown in Figures \ref{f:fig1} and \ref{f:fig2} remain roughly the same regardless of whether or not we apply model-dependent progenitor hydrogen-envelope mass and radius constraints. %apart from a slight decrease in the peak in the distribution of pre-SN orbital period around $\logten (P_{\rm orb,preSN}/$d$) \sim 3.8$ for $\epsilon$ = 0.1. 
%Models run without MLT++ favor slightly smaller pre-SN progenitor envelope masses, slightly larger helium-core masses, and slightly larger radii: the dominant peaks are at $M_{\rm H~env,preSN,1} \sim 0.1M_\sun$, $\logten (M_{\rm He~core,preSN,1}/M_\sun) \sim 0.8$, and $R_{\rm preSN,1} \sim 100R_\sun$, respectively. The distribution of remaining properties are robust to the effects of MLT++.

In figure \ref{f:fig3} we show the distribution of locations on the Hertzsprung-Russell (H-R) diagram of potential single and binary star progenitors of SN 2016gkg. The luminosities of binary progenitors are smaller than of single-star progenitors. This is a consequence of smaller pre-SN helium-core masses for binary progenitors (see Figure \ref{f:fig2}). The secondaries of binary progenitors mostly lie on the main-sequence and are less luminous than their primaries with $M_{\rm bol} \gtrsim -8.5$. Some binary progenitors with initial mass ratios $\sim 1$ have evolved secondaries that are on the red-giant branch (RGB). 
We find that it is unlikely for the secondary to lie between the main sequence and the RGB, which is the case for the companion of SN 1993J \citep{2004Natur.427..129M}. Otherwise, no strong constraints can be placed on binary companions' location on the H-R diagram. We also note that in binaries with secondaries of luminosities similar to that of the primary, flux from the secondary may contaminate flux from the primary, making our calculations inconsistent with the derivation of observed constraints. We find that the total posterior probability of progenitors having secondaries with luminosities within a factor of 2 of the primary (and thus potentially contaminated) is $\sim 12$\% for $\epsilon = 0.1$ and $\sim 4$\% for $\epsilon = 0.5$.

X-ray/radio observations can be used to infer the CSM density around SN progenitors and thus trace the mass loss history of the progenitor star. We use our models to infer the CSM density at $10^{16}\,$cm to compare with the results of \citet{2017arXiv170405865M} (example Figure 6). Our models have a SN Ibc-like mass loss history: $v_{\rm wind} \sim 50 - 250$ km s$^{-1}$ and $\dot{M}_{\rm wind} \sim 10^{-4.8} - 10^{-6} M_\sun$ yr$^{-1}$. This is because all potential binary progenitors detach before core-collapse. %We get consistent results with our models run without MLT++. 
If future measurements indicate that SN 2016gkg also experienced high mass-loss rates ($\sim 10^{-4} M_\sun$ yr$^{-1}$), like those for other Type IIb SNe in the aforementioned study, then it would indicate that the progenitor experienced a period of enhanced mass loss just before explosion.

\renewcommand{\arraystretch}{1.5}
\begin{deluxetable}{cccC{1.1cm}C{1.1cm}C{1.1cm}C{1.1cm}}
\centering
\tablecaption{{Probability of a binary star progenitor of SN 2016gkg} \label{t:rates}}
\tablehead{
\colhead{$\alpha$} & \colhead{$\beta$} & \colhead{$\gamma$} & \multicolumn{2}{c}{$P_{\rm binary}|L,T_{\rm eff}$} & \multicolumn{2}{c}{$P_{\rm binary}|L,T_{\rm eff},M_{\rm env},R$} \\
\colhead{} & \colhead{} & \colhead{} & \colhead{$\epsilon = 0.1$} & \colhead{$\epsilon = 0.5$} & \colhead{$\epsilon = 0.1$} & {$\epsilon = 0.5$}}
\startdata
 				& \multirow{2}{*}{-2.0}	& 0.00	& 0.13 & 0.08 & 0.28 & 0.17 \\
 				& 				 	& -0.22	& 0.11 & 0.06 & 0.25 & 0.15 \\
				\cline{2-7}
\multirow{2}{*}{1.6}	& \multirow{2}{*}{-1.0} 	& 0.00 	& 0.21 & 0.12 & 0.42 & 0.26 \\
 				& 				 	& -0.22 	& 0.18 & 0.11 & 0.38 & 0.23 \\
				\cline{2-7}
			 	& \multirow{2}{*}{0.0} 	& 0.00	& 0.29 & 0.17 & 0.52 & 0.34 \\
 				& 				 	& -0.22	& 0.25 & 0.15 & 0.48 & 0.30 \\
\hline 
 				& \multirow{2}{*}{-2.0} 	& 0.00	& 0.16 & 0.10 & 0.34 & 0.22 \\
 				& 					& -0.22	& 0.14 & 0.08 & 0.31 & 0.19 \\
				\cline{2-7}
\multirow{2}{*}{2.3} 	& \multirow{2}{*}{-1.0}	& 0.00 	& 0.25 & 0.15 & 0.48 & 0.32 \\
 				& 				 	& -0.22 	& 0.22 & 0.13 & 0.44 & 0.28 \\
				\cline{2-7}
				& \multirow{2}{*}{0.0} 	& 0.00	& 0.34 & 0.21 & 0.59 & 0.40 \\
 				&				 	& -0.22	& 0.31 & 0.18 & 0.55 & 0.36 \\
\hline
 				& \multirow{2}{*}{-2.0} 	& 0.00	& 0.20 & 0.12 & 0.41 & 0.27 \\
 				& 				 	& -0.22	& 0.17 & 0.11 & 0.37 & 0.24 \\
				\cline{2-7}
\multirow{2}{*}{3.0} 	& \multirow{2}{*}{-1.0} 	& 0.00 	& 0.31 & 0.19 & 0.56 & 0.38 \\
 				& 				 	& -0.22 	& 0.28 & 0.17 & 0.52 & 0.35 \\
				\cline{2-7}
				& \multirow{2}{*}{0.0} 	& 0.00	& 0.41 & 0.25 & 0.66 & 0.47 \\
 				& 					& -0.22	& 0.37 & 0.22 & 0.62 & 0.43 
\enddata
\tablecomments{$f_{\rm bin} = 0.5$, and $\alpha$, $\beta$, and $\gamma$ are parameters for the priors on the initial mass, $\logten M_{\rm ZAMS}$, initial mass ratio, $q_{\rm ZAMS}$, and initial orbital period, $P_{\rm orb}$, respectively (see Eqs. \ref{e:IMF}, \ref{e:f(q)}, and \ref{e:f(P)}).}
\end{deluxetable}

Finally, we compute the probability that the progenitor of SN 2016gkg was a binary: the total posterior probability of all model binary star progenitors divided by total posterior probability of all model single and binary star progenitors. In Table \ref{t:rates} we list probabilities of a binary star progenitor of SN 2016gkg not applying and applying model-dependent progenitor hydrogen-envelope mass and radius constraints, for $f_{\rm bin} = 0.50$ and various values of $\alpha$, $\beta$, and $\gamma$. 
We find that the probability of a binary star progenitor of SN 2016gkg with $\epsilon$ =  0.1 and 0.5 not-given (given) progenitor hydrogen-envelope mass and radius constraints is 22\% (44\%) and 13\% (28\%, respectively, for our fiducial values of $\alpha$ (2.3), $\beta$ (-1.0), and $\gamma$ (-0.22).

\section{Conclusions} \label{s:conclusions}

We use Bayesian inference and a large grid of single and binary star models to derive the distributions of potential progenitors and companions of SN 2016gkg. We find that potential binary star progenitors have lower initial primary mass and pre-SN hydrogen-envelope and helium-core mass than single-star progenitors. 
The probability that the progenitor of SN 2016gkg was a binary with $\epsilon$ = 0.1 (0.5) is 22\% (13\%) if we only use luminosity and effective temperature constraints on the progenitor star. Applying model-dependent observational constraints on the progenitor hydrogen-envelope mass and radius rule out several single-star progenitors, favoring a binary as the progenitor of SN 2016gkg (44\% for $\epsilon$ = 0.1 and 28\% for $\epsilon$ = 0.5). In either case, there is no clear preference for a binary star progenitor of SN 2016gkg. This is in contrast to binaries being the currently favored progenitors of Type IIb SNe. We find that, a binary companion, if present, has $M_{\rm bol} \gtrsim -8.5$ and is a main-sequence or red-giant star. As such, we are unable to find strong constraints on the nature of the companion star. Constraints on the progenitor helium-core mass can help tighten constraints on the progenitor. Similarly, improved constraints on the progenitor luminosity can significantly narrow the parameter space for progenitors. We would like to stress that the parameter space for Type IIb SN progenitors is strongly dependent on the progenitor metallicity. At lower metallicities, the parameter space for binary progenitors of Type IIb SNe widens significantly \citep{2017ApJ...840...10Y}. We expect that the results presented here will differ strongly at low metallicities with the probability of a binary progenitor increasing significantly. Nevertheless, the probability of a binary progenitor derived at solar metallicity represents a lower limit. At lower metallicities, the binary star channel towards Type IIb SNe dominates for a couple of reasons. First, the parameter space for binary star Type IIb SN progenitors widens significantly at lower metallicity \citep{2017ApJ...840...10Y}. Second, single star Type IIb SN progenitors are produced at higher masses (due to the scaling in wind mass-loss) and are thus disfavored by the IMF. A detailed investigation into the effects of metallicity on the relative importance of single and binary progenitors of Type IIb SNe would be an interesting line of future investigation.

\acknowledgments
NS, PM, and VK acknowledge support from NSF grant AST-1517753. NS acknowledges support from NSF grant DGE-1450006 and from Northwestern University. This work was performed in part at the Aspen Center for Physics, which is supported by National Science Foundation through grant PHY-1607611.
%This research was partially supported by the National Science Foundation through grant -Enter Award Number here-.
This research was supported by computational resources provided for the Quest high performance computing facility at Northwestern University which is jointly supported by the National Science Foundation through grant NSF PHY-1126812, the Office of the Provost, the Office for Research, and Northwestern University Information Technology.

\end{document}